\hfuzz 2pt
\font\titlefont=cmbx10 scaled\magstep1
\font\bigrm=cmr10 scaled\magstep1
\font\bigmath=cmmi10 scaled\magstep2
\magnification=\magstep1

\null
\vskip 1.5cm
\centerline{ 
{\bigmath \char'042}{\bigrm \char'023}\hskip -2pt
{\bigmath \char'075\char'042} \ \
{\titlefont VIOLATES BELL'S INEQUALITY}}
\vskip 3.5cm
\centerline{\bf F. Benatti}
\smallskip
\centerline{Dipartimento di Fisica Teorica, Universit\`a di Trieste}
\centerline{Strada Costiera 11, 34014 Trieste, Italy}
\centerline{and}
\centerline{Istituto Nazionale di Fisica Nucleare, Sezione di 
Trieste}
\vskip 1cm
\centerline{\bf R. Floreanini}
\smallskip
\centerline{Istituto Nazionale di Fisica Nucleare, Sezione di 
Trieste}
\centerline{Dipartimento di Fisica Teorica, Universit\`a di Trieste}
\centerline{Strada Costiera 11, 34014 Trieste, Italy}
\vskip 3cm
\centerline{\bf Abstract}
\smallskip
\midinsert
\narrower\narrower\noindent
We formulate a set of Bell's inequalities for the system of two correlated
neutral kaons coming from the decay of a $\phi$-meson, without assuming
$CP$ and $CPT$ invariance. We show that a non-vanishing value of the
phenomenological parameter $\varepsilon'$ would violate such inequalities,
ruling out Bell's locality.
\endinsert
\bigskip
\vfil\eject

The neutral kaon system has proven to be one of the most useful laboratory for
studying many aspects of modern particle physics. Indeed,
the study of the phenomena associated to kaon production, propagation and decay
has provided strong experimental confirmations for various predictions 
of the Standard Model.

Recently, the $K^0$-$\overline{K^0}$ system 
has also been proposed as the natural
system to look for possible new phenomena leading to loss of quantum coherence
and $CPT$ violating effects. [1-3] These arise as natural consequences of 
effective completely positive dynamics for the kaon system and could be of
particular relevance in the study of correlated kaons. [4-6]

When a $\phi$-meson at rest decays into two neutral kaons, the final state has
the property of being antisymmetric in the spatial part, due to angular momentum
conservation. Although the two kaons fly apart with opposite momenta, they
remain quantum mechanically correlated in a way that is very similar to the
entanglement of two spin-1/2 particles in a singlet state. Therefore, as in the
spin case, by studying the evolution of certain observables of the correlated
kaons state, one can perform fundamental tests on the 
behaviour of entangled systems.

In the following we shall focus on the property  
called Bell's locality [7-10] and discuss
whether, by looking at experimentally observable quantities, 
the system of the two neutral kaons coming from the $\phi$-decay
satisfies this condition. Typically, observables that can be studied
in the two-kaon system are the probabilities 
${\cal P}(f_1,\tau_1;f_2,\tau_2)$ that one kaon decays into the final state
$f_1$ at proper time $\tau_1$, while the other kaon decays into the final
state $f_2$ at proper time $\tau_2$. The requirement of Bell's locality can be
translated into certain inequalities that the probabilities $\cal P$ should
satisfy (Bell's inequalities).

We shall see that in the standard quantum mechanical description of the
correlated two-kaon system, these inequalities are violated by the
phenomenological constant $\varepsilon'$, 
that parametrizes small direct $CP$ and $CPT$
violating effects in the decays of the kaons in two pions. Therefore,
if in an actual experiment the parameter $\varepsilon'$ is found to be
nonzero, then Bell's locality will be ruled out.

\bigskip

For our discussion, we shall adopt a particle description of the phenomena
related to the time evolution and decay of the $K^0$-$\overline{K^0}$ system.
This means that the transition probabilities we shall be dealing with are
actually $S$-matrix elements, {\it i.e.} matrix elements of the scattering
operator $S$ (or better, the transition operator $T=1-S$) between asymptotic
particle states. This description is standard in particle physics since it
is the closest to the actual experimental situations in which the various decay
products are identified in physical detectors. 
The $K^0$-$\overline{K^0}$ system
can then be effectively described by means 
of a two-dimensional Hilbert space. [11]
A useful orthonormal basis in this space is given by the $CP$-eigenstates 
$|K_1\rangle$ and $|K_2\rangle$:
$$
|K_1\rangle={1\over\sqrt{2}}\Big[|K^0\rangle+|\overline{K^0}\rangle\Big]\ ,\quad
|K_2\rangle={1\over\sqrt{2}}\Big[|K^0\rangle-|\overline{K^0}\rangle\Big]\ .
\eqno(1)
$$
For the discussion that follows, we find it convenient to describe
the states of a physical system by means of density matrices.
These are hermitian matrices $\rho$, with positive eigenvalues
and unit trace. 
In the case of the kaon system, $\rho$ is a two-dimensional
matrix. With respect to the basis (1) it can be written as:
$$
\rho=\left(\matrix{
\rho_1&\rho_3\cr
\rho_4&\rho_2}\right)\ , \eqno(2)
$$
where $\rho_4\equiv\rho_3^*$, and $*$ signifies complex conjugation.

The evolution in time of this matrix is described by the following equation
$$
\eqalignno{
&{\partial\rho(t)\over\partial t}=-iH\, \rho(t)+i\rho(t)\, H^\dagger\ , &(3a)\cr
&\rho \mapsto \rho(t)\equiv\gamma_t[\rho]=e^{-itH}\, \rho\, e^{itH^\dagger}\ .
&(3b)}
$$
Since the kaon system is unstable, the effective hamiltonian $H$
includes a nonhermitian part that characterizes 
the natural width of the states:
$$
H=M-{i\over 2}{\mit\Gamma}\ ,\eqno(4)
$$ 
with $M$ and $\mit\Gamma$ positive hermitian $2\times 2$ matrices.
The entries of these matrices can be expressed in terms of the 
complex parameters $\epsilon_S$, $\epsilon_L$, appearing in the
eigenstates of $H$, 
$$
|K_S\rangle=N_S\left(\matrix{1\cr\epsilon_S}\right)\ ,\quad
|K_L\rangle=N_L\left(\matrix{\epsilon_L\cr 1}\right)\ ,\eqno(5)
$$
and the four real parameters, $m_S$, $\gamma_S$ and $m_L$, $\gamma_L$
characterizing the eigenvalues of $H$: 
$$
\lambda_S=m_S-{i\over 2}\gamma_S\ ,\quad
\lambda_L=m_L-{i\over 2}\gamma_L\ ,\eqno(6)
$$
where $N_S=(1+|\epsilon_S|^2)^{-1/2}$ and $N_L=(1+|\epsilon_L|^2)^{-1/2}$ 
are normalization factors; $\gamma_S$, $\gamma_L$ and $m_S$, $m_L$
are the physical decay widths and masses of the
states $K_S$ and $K_L$. Using (5), these states are described by the following
two density matrices:
$$
\eqalignno{
&\rho_L=|N_L|^2\left(\matrix{|\epsilon_L|^2 & \epsilon_L\cr
                            \epsilon_L^* & 1}\right)\ , &(7a) \cr
&\rho_S=|N_S|^2\left(\matrix{1 & \epsilon_S^*\cr
                            \epsilon_S & |\epsilon_S|^2}\right)\ . &(7b)}
$$
The constants $\epsilon_S$ and $\epsilon_L$ parametrize the so-called
indirect $CP$ and (for $\epsilon_S\neq\epsilon_L$) $CPT$ violating effects;
if we ignore these small effects, $\rho_L$ and $\rho_S$ would describe
the $CP$-eigenstates in (1).

The time evolution of a system of two correlated neutral kaons, 
as those coming from the decay of a $\phi$-meson, can be derived
from the single-kaon dynamical map $\gamma_t$ in $(3b)$. 
First note that, since the $\phi$-meson has spin 1, 
its decay into two spinless bosons produces
an antisymmetric spatial state. In the $\phi$ rest frame, the two neutral kaons
are produced flying apart with opposite momenta; in the basis $|K_1\rangle$,
$|K_2\rangle$, the resulting state can be described by:
$$
|\psi_A\rangle= {1\over\sqrt2}\Big(|K_1,-p\rangle \otimes  |K_2,p\rangle -
|K_2,-p\rangle \otimes  |K_1,p\rangle\Big)\ .\eqno(8)
$$
The corresponding density operator $\rho_A$ is a $4\times 4$ matrix
that can be conveniently written in terms of single kaon projectors:
$$
P_1\equiv |K_1\rangle\langle K_1|=
    \left(\matrix{1 & 0\cr 0 & 0\cr}\right)\ ,\qquad
P_2\equiv |K_2\rangle\langle K_2|=
    \left(\matrix{0 & 0\cr 0 & 1\cr}\right)\ ,\eqno(9a)
$$
and the off-diagonal operators
$$
P_3\equiv |K_1\rangle\langle K_2|=
    \left(\matrix{0 & 1\cr 0 & 0\cr}\right)\ ,\qquad
P_4\equiv |K_2\rangle\langle K_1|=
    \left(\matrix{0 & 0\cr 1 & 0\cr}\right)\ .\eqno(9b)
$$
Explicitly, one finds:
$$
\rho_A={1\over 2}\Big[P_1\otimes P_2\ +\ P_2\otimes P_1\ -\ 
P_3\otimes P_4\ -\ P_4\otimes P_3\Big]\ .\eqno(10)
$$
Once produced in a $\phi$ decay, the kaons evolve 
independently in time each according to the
map $\gamma_t$ in $(3b)$. The resulting total evolution map 
$\rho_A\mapsto \Gamma_t[\rho_A]$ is therefore of factorized form
$\Gamma_t=\gamma_t \otimes \gamma_t$. More explicitly, 
the density matrix that describes a situation in which
the first kaon has evolved up to proper time $\tau_1$ 
and the second up to proper time $\tau_2$ is given by:
$$
\eqalign{
\rho_A(\tau_1&,\tau_2)\equiv
\big(\gamma_{\tau_1}\otimes\gamma_{\tau_2}\big)\big[\rho_A\big]\cr
=&{1\over 2}\Big[P_1(\tau_1)\otimes P_2(\tau_2)\ 
+\ P_2(\tau_1)\otimes P_1(\tau_2)\ 
- P_3(\tau_1)\otimes P_4(\tau_2)-P_4(\tau_1)\otimes P_3(\tau_2)\Big]\ ,}
\eqno(11)
$$
where $P_i(\tau_1)$ and $P_i(\tau_2)$, $i=1,2,3,4$, represent the evolution
according to (3) of the initial operators (9), up to the time $\tau_1$
and $\tau_2$, respectively:
$$
P_i(\tau)=e^{-i\tau\, H}\, P_i\, e^{i\tau H^\dagger}\ ,\quad i=1,2,3,4\ .
\eqno(12)
$$

The formula (11) can now be used to compute the time evolution of characteristic
observables for the two-kaon system; indeed, any physical property of this
system can be extracted from the density matrix $\rho_A(\tau_1,\tau_2)$
by taking its trace with suitable hermitian operators. The typical
observables that can be studied are double
decay probabilities ${\cal P}(f_1,\tau_1; f_2,\tau_2)$, 
{\it i.e.} the probabilities that one kaon decays
into a final state $f_1$ at proper time $\tau_1$, while the other kaon 
decays into the final state $f_2$ at proper time $\tau_2$. [12]
Using (11), one explicitly finds:
$$
\eqalign{
{\cal P}(f_1,\tau_1;& f_2,\tau_2)\equiv 
\hbox{Tr}\Big[\Big({\cal O}_{f_1}\otimes{\cal O}_{f_2}\Big) 
\rho_A(\tau_1,\tau_2)\Big]\cr
=&{1\over 2}\Big[
\hbox{Tr}\big\{P_1(\tau_1)\,{\cal O}_{f_1}\big\}\
\hbox{Tr}\big\{P_2(\tau_2)\,{\cal O}_{f_2}\big\}\ +\
\hbox{Tr}\big\{P_2(\tau_1)\,{\cal O}_{f_1}\big\}\ 
\hbox{Tr}\big\{P_1(\tau_2)\,{\cal O}_{f_2}\big\}\cr
-&\hbox{Tr}\big\{P_3(\tau_1)\,{\cal O}_{f_1}\big\}\
\hbox{Tr}\big\{P_4(\tau_2)\,{\cal O}_{f_2}\big\}\ -\
\hbox{Tr}\big\{P_4(\tau_1)\,{\cal O}_{f_1}\big\}\
\hbox{Tr}\big\{P_3(\tau_2)\,{\cal O}_{f_2}\big\}\Big]\ ,
}\eqno(13)
$$
where ${\cal O}_{f_1}$ and 
${\cal O}_{f_2}$ represent the
$2\times 2$ projector matrices, 
describing the decay of a single kaon into the
final states $f_1$ and $f_2$, respectively.

Useful observables are associated with the decays of the neutral
kaons into two pions and into semileptonic states. We shall be as
general as possible and use matrices ${\cal O}_f$ that encode possible
$CP$ and $CPT$ violating effects also in the decay amplitudes.

The amplitudes for the decay of a $K^0$ state into $\pi^+\pi^-$
and $\pi^0\pi^0$ final states are usually parametrized as follows,
in terms of the $s$-wave phase-shifts $\delta_i$ and the complex coefficients
$A_i$, $B_i$, $i=1,\ 2$: [13]
$$
\eqalignno{
&{\cal A}(K^0\rightarrow \pi^+\pi^-)=(A_0+B_0)\, e^{i\delta_0}+{1\over\sqrt2}\,
(A_2+B_2)\, e^{i\delta_2}\ ,&(14a)\cr
&{\cal A}(K^0\rightarrow \pi^0\pi^0)=(A_0+B_0)\, e^{i\delta_0}-\sqrt{2}\,
(A_2+B_2)\, e^{i\delta_2}\ ,&(14b)}
$$
where the indices 0, 2 refers to the total isospin $I$. The amplitudes for
the $\overline{K^0}$ decays are obtained from these with the substitutions:
$A_i\rightarrow A_i^*$ and $B_i\rightarrow -B_i^*$.
The imaginary parts of $A_i$ signals direct $CP$-violation, while a non zero
value for $B_i$ will also break $CPT$ invariance.

To construct the operators that describe these two pion decays in
the formalism of density matrices, one has to pass to the $K_1$, $K_2$
basis of Eq.(1). It is convenient to label the corresponding decay amplitudes
as follows:
$$
\eqalign{
&{\cal A}(K_1\rightarrow \pi^+\pi^-)=X_{+-}\ ,\cr
&{\cal A}(K_1\rightarrow \pi^0\pi^0)=X_{00}\ ,}\qquad\quad
\eqalign{
&{\cal A}(K_2\rightarrow \pi^+\pi^-)=Y_{+-}\ 
{\cal A}(K_1\rightarrow \pi^+\pi^-)\ ,\cr
&{\cal A}(K_2\rightarrow \pi^0\pi^0)=Y_{00}\ 
{\cal A}(K_1\rightarrow \pi^0\pi^0)\ .}
\eqno(15)
$$
The complex parameters $X$ and $Y$ can be easily expressed in terms
of $A_i$, $B_i$ and $\delta_i$ of (14).
For the dominant amplitudes, one easily finds:
$$
\eqalignno{
&X_{+-}=\sqrt{2}\, \Big[{\cal R}e(A_0)+i{\cal I}m(B_0)\Big]\, e^{i\delta_0}
+\Big[{\cal R}e(A_2)+i{\cal I}m(B_2)\Big]\, e^{i\delta_2}\ ,&(16a)\cr
&X_{00}=\sqrt{2}\, \Big[{\cal R}e(A_0)+i{\cal I}m(B_0)\Big]\, e^{i\delta_0}
-2\Big[{\cal R}e(A_2)+i{\cal I}m(B_2)\Big]\, e^{i\delta_2}\ .&(16a)}
$$
On the other hand, the $K_2$ amplitudes are suppressed, since they involve only
$CP$ and $CPT$ violating terms. For the considerations that follow,
it will be sufficient to work in the approximation that keeps only the
dominant terms in these violating parameters. Then, the amplitude ratios 
$Y_{+-}$ and $Y_{00}$ can be written as:
$$
Y_{+-}=\varepsilon-\epsilon_L+\varepsilon^\prime\ ,\qquad
Y_{00}=\varepsilon-\epsilon_L-2\varepsilon^\prime\ ,
\eqno(17)
$$
where the parameters $\varepsilon$ and $\varepsilon'$ take the familiar
expressions:
$$
\varepsilon=\bigg[{\epsilon_L+\epsilon_S\over2} +
i\, {{\cal I}m(A_0)\over {\cal R}e(A_0)}\bigg]+
\bigg[{\epsilon_L-\epsilon_S\over2} +
{{\cal R}e(B_0)\over {\cal R}e(A_0)}\bigg]\ ,\eqno(18)
$$
and
$$
\varepsilon^\prime= {i e^{i(\delta_2-\delta_0)}\over\sqrt2}\,
{{\cal R}e(A_2)\over {\cal R}e(A_0)}\, \bigg[
{{\cal I}m(A_2)\over {\cal R}e(A_2) }
-{{\cal I}m(A_0)\over {\cal R}e(A_0) }\bigg]
+i\bigg[{{\cal R}e(B_0)\over  {\cal R}e(A_0) }
-{ {\cal R}e(B_2)\over {\cal R}e(A_0)}\bigg]\ .\eqno(19)
$$
The first, second square brackets in (18) and (19) contain the $CP$,
respectively $CPT$, violating parameters arising from the neutral kaon mass and
decay matrices, as well as from their decay amplitudes. The factor
$\omega={\cal R}e(A_2)/{\cal R}e(A_0)$ corresponds to the suppression due to
the $\Delta I=1/2$ rule in the kaon two-pion decays. This ratio is known to be
small; therefore, in presenting the above formulas, first order terms in the
small parameters multiplied by $\omega^2$ have been consistently neglected.
Note that the expressions for $\varepsilon$ and $\varepsilon'$ given in
(18) and (19) are independent from the phase conventions adopted in describing
the kaon states.

Using (15), the operators that describe the 
$\pi^+\pi^-$ and $\pi^0\pi^0$ final states and include direct $CP$
and $CPT$ violations are readily found: [3, 5]
$$
\widetilde{\cal O}_{+-}=|X_{+-}|^2\ \left[\matrix{1&Y_{+-}\cr
                              Y_{+-}^*&|Y_{+-}|^2\cr}\right]\ ,\qquad
\widetilde{\cal O}_{00}=|X_{00}|^2\ \left[\matrix{1&Y_{00}\cr
                              Y_{00}^*&|Y_{00}|^2\cr}\right]\ .
\eqno(20)
$$
One can check this result by computing the corresponding 
decay rates for the physical states $K_L$ and $K_S$. These are simply given by
the traces of (20) with the density matrices in (7):
$$
\eqalignno{
&\big|{\cal A}(K_L\rightarrow \pi^+\pi^-)\big|^2\equiv
{\rm Tr}\Big(\widetilde {\cal O}_{+-}\, \rho_L\Big)=
|X_{+-}|^2\, |N_L|^2\, |\epsilon_L+Y_{+-}|^2\ , &(21a)  \cr
&\big|{\cal A}(K_S\rightarrow \pi^+\pi^-)\big|^2\equiv
{\rm Tr}\Big(\widetilde {\cal O}_{+-}\, \rho_S\Big)=  
|X_{+-}|^2\, |N_S|^2\, |1+\epsilon_S\, Y_{+-}|^2\ . &(21b)}
$$
From this result, one recovers, to leading order in $CP$ and $CPT$
violation and $\Delta I=1/2$ enhancement, the familiar result:
$$
|\eta_{+-}|^2\equiv\left|{{\cal A}(K_L\rightarrow \pi^+\pi^-)\over
{\cal A}(K_S\rightarrow \pi^+\pi^-)}\right|^2\simeq
|\varepsilon+\varepsilon'|^2\ .\eqno(22)
$$
In an analogous way, for the $2\pi^0$ final state one finds
$$
|\eta_{00}|^2\equiv\left|{{\cal A}(K_L\rightarrow 2\pi^0)\over
{\cal A}(K_S\rightarrow 2\pi^0)}\right|^2\simeq
|\varepsilon-2\varepsilon'|^2\ .\eqno(23)
$$
The phenomenological parameters $\varepsilon$ and $\varepsilon'$ that 
parametrize $\eta_{+-}$ and $\eta_{00}$, the $K_L$, $K_S$ amplitudes
ratio for the $\pi^+\pi^-$ and $2\pi^0$ decays, are directly accessible
to experiments.

It is important to notice that unlike the matrices ${\cal O}_f$ in (13),
the matrices in (20) are not normalized since they give rise
decay rates. Indeed, if we call $T$ the transition matrix responsible
for the decay of the $K_L$ state into the final $\pi^+\pi^-$ asymptotic state,
then one has
$$
{\rm Tr}\Big(\widetilde {\cal O}_{+-}\, \rho_L\Big)
\equiv\Big|\langle \pi^+\pi^-|\ T\ | K_L\rangle\Big|^2\ ;\eqno(24)
$$
equivalently, the $2\times 2$ hermitian matrix $\widetilde{\cal O}_{+-}$ can be written as
$$
\widetilde{\cal O}_{+-}=T^\dagger\ |\pi^+\pi^-\rangle\langle \pi^+\pi^-|\ T\ .
\eqno(25)
$$
In other words, it is the operator $T$ that allows an effective two-dimensional
description of the projector $|\pi^+\pi^-\rangle\langle \pi^+\pi^-|$,
which acts on the Hilbert space of the decay products.

However, we are interested in computing decay probabilities
and not decay rates. Therefore, in the expression (13) for the joint
probabilities $\cal P$, normalized 
expressions for the observable ${\cal O}_{+-}$ 
must be used. These are obtained by dividing the expression (25) 
by the norm of the vector $T^\dagger |\pi^+\pi^-\rangle$,
or more simply by dividing the expressions in (20) 
by their corresponding traces. Explicitly,
for the normalized two-pion observables one finds the projectors:
$$
{\cal O}_{+-}=N_{+-}\ \left[\matrix{1&Y_{+-}\cr
                              Y_{+-}^*&|Y_{+-}|^2\cr}\right]\ ,\qquad
{\cal O}_{00}=N_{00}\ \left[\matrix{1&Y_{00}\cr
                              Y_{00}^*&|Y_{00}|^2\cr}\right]\ ,\eqno(26)
$$
where $N_{+-}=(1+|Y_{+-}|^2)^{-1}$ and $N_{00}=(1+|Y_{00}|^2)^{-1}$.

In a similar way one can derive the expressions for the observables
that describe the decay of neutral kaons into the semileptonic states
$\pi^-\ell^+\nu$ and $\pi^+\ell^-\bar\nu$. The amplitudes for
the decay of a $K^0$ or a $\overline{K^0}$ into these final states
are usually parametrized by three complex
constants $x$, $y$ and $z$  as follows: [14]
$$
\eqalignno{
&{\cal A}(K^0\rightarrow\pi^-\ell^+\nu)={\cal M} (1-y)\ , &(27a)\cr
&{\cal A}(\overline{K^0}\rightarrow\pi^+\ell^-\bar\nu)=
{\cal M}^* (1+y^*)\ , &(27b)\cr
&{\cal A}(K^0\rightarrow\pi^+\ell^-\bar\nu)= z\, 
{\cal A}(\overline{K^0}\rightarrow\pi^+\ell^-\bar\nu)\ , &(27c)\cr
&{\cal A}(\overline{K^0}\rightarrow\pi^-\ell^+\nu)=
x\, {\cal A}(K^0\rightarrow\pi^-\ell^+\nu)\ , &(27d) }
$$
where ${\cal M}$ is a common factor.
The $\Delta S=\Delta Q$ rule would forbid the decays
$K^0\rightarrow\pi^+\ell^-\bar\nu$ and 
$\overline{K^0}\rightarrow\pi^-\ell^+\nu$, so that the parameters $x$ and $z$
measure the violations of this rule ($\Delta S$ and $\Delta Q$ 
represent the difference in strangeness and electric charge between the final 
and initial hadronic particles). Instead, $CPT$-invariance would require $y=\,0$.

From the parametrization in (27), one can deduce the decay amplitudes
from the states (1) of definite $CP$, and therefore the following
expressions for the two operators describing the semileptonic decay rates: [3, 5]
$$
\eqalignno{
&\widetilde{\cal O}_{\ell^+}={|{\cal M}|^2\over2}\,
|1-y|^2\ \left[\matrix{|1+x|^2&(1+x^*)(1-x)\cr
                                (1+x)(1-x^*)&|1-x|^2\cr}\right]\ ,&(28a)\cr
&\widetilde{\cal O}_{\ell^-}={|{\cal M}|^2\over2}\,
|1+y|^2\ \left[\matrix{|z+1|^2&(z^*+1)(z-1)\cr
                                (z+1)(z^*-1)&|z-1|^2\cr}\right]\ .&(28b)}
$$
As in the case of the decay into two pions, the
observables ${\cal O}_{\ell^+}$ and ${\cal O}_{\ell^-}$ useful for computing
decay probabilities, can be obtained from (28) by dividing these expressions
by the corresponding traces.

Once inserted in the general formula (13), the four observables 
${\cal O}_{+-}$, ${\cal O}_{00}$, 
${\cal O}_{\ell^+}$ and ${\cal O}_{\ell^-}$ allow determining various joint
probabilities for the system of two kaons
coming from the decay of a $\phi$-meson. These double probabilities enter
a class of inequalities that can be derived 
from the hypothesis of Bell's locality.
Before giving explicit expressions for those probabilities, we shall
derive and discuss these relations (Bell's inequalities).

In actual experimental setups (the so called $\phi$-factories),
one studies the decay of a $\phi$-meson into two neutral kaons by counting
the occurrence of the various final decay states for the two kaons.
For instance, the probability ${\cal P}(f_1,\tau_1; -,\tau_2)$ of finding
a certain final state $f_1$ at proper time $\tau_1$ for one of the two kaons
and any decay mode for the second one at proper time $\tau_2$
can be experimentally obtained as the ratio 
of partial $(f_1,\tau_1; -,\tau_2)$-counts over the total $\phi$-decays. 
A similar argument holds for the double
probability ${\cal P}(f_1,\tau_1;f_2,\tau_2)$.

Let us now suppose that we can describe the system of the two kaons 
by a set of variables that we globally call $\lambda$.
We assume that the description in terms of the set $\lambda$ is the best
available characterization of the system. In particular, given the variables
$\lambda$, we expect a well-defined probability $p_\lambda(f_1,\tau_1;-,\tau_2)$
of detecting a final decay state $(f_1,\tau_1)$ for one of the two kaons
and any state $(-,\tau_2)$ for the second one,
and a probability $p_\lambda(f_1,\tau_1;f_2,\tau_2)$ of detecting a final
decay state $(f_1,\tau_1)$ for one kaon and a final decay state
$(f_2,\tau_2)$ for the second one. The average probabilities
${\cal P}(f_1,\tau_1; -;\tau_2)$ and ${\cal P}(f_1,\tau_1;f_2,\tau_2)$ 
are then given by
$$
\eqalignno{
&{\cal P}(f_1,\tau_1; -,\tau_2)=\int d\lambda\, \rho(\lambda)\, 
p_\lambda(f_1,\tau_1;-, \tau_2)\ , &(29a)\cr
&{\cal P}(f_1,\tau_1;f_2,\tau_2)=\int d\lambda\, \rho(\lambda)\,
p_\lambda(f_1,\tau_1;f_2,\tau_2)\ , &(29b)}
$$
where $\rho(\lambda)$ is a normalized probability density characterizing
the ensemble of initial $\phi$ particles. It should be stressed that this
description of the $\phi$-decay into two neutral kaons is rather general,
and surely can be made to agree with ordinary quantum mechanics.

Since the decays of the two kaons coming from a $\phi$-meson are localized
events, usually very well spatially separated, it seems natural to assume
that the probabilities\hfill\break
$p_\lambda(f_1,\tau_1;-,\tau_2)$ 
and $p_\lambda(-,\tau_1;f_2,\tau_2)$ are independent. 
In other words, one is led to consider a description of
the two-kaon system for which:
$$
p_\lambda(f_1,\tau_1;f_2,\tau_2)=p_\lambda(f_1,\tau_1;-,\tau_2)\ 
p_\lambda(-,\tau_1;f_2,\tau_2)\ .\eqno(30)
$$
This condition is known as Bell's locality condition. Measurable consequences
of this assumption can be easily derived. 
We will now adapt the arguments developed in Ref.[8] for stationary spin
singlets to the case of time-evolving correlated kaon systems.

By noting that for any set of four positive numbers $x_1$, $x_2$, $x_3$, $x_4$
less or equal to 1 the following inequality holds:
$$
x_1\, x_2-x_1\, x_4+x_3\, x_2+x_3\, x_4\leq x_3+x_2\ ,\eqno(31)
$$
using (30) one easily deduce the following relation among double probabilities:
$$
\eqalign{
p_\lambda(f_1,\tau_1;f_2,\tau_2)-p_\lambda(f_1,\tau_1;f_4,\tau_2)
+p_\lambda &(f_3,\tau_1;f_2,\tau_2)+p_\lambda(f_3,\tau_1;f_4,\tau_2)\cr
&\leq p_\lambda(f_3,\tau_1;-,\tau_2)+p_\lambda(-,\tau_1;f_2,\tau_2)\ .}\eqno(32)
$$
Integrating over $\lambda$ with weight $\rho(\lambda)$, one finally obtains
$$
\eqalign{
{\cal P}(f_1,\tau_1;f_2,\tau_2)-{\cal P}(f_1,\tau_1;f_4,\tau_2)
+{\cal P}&(f_3,\tau_1;f_2,\tau_2)+{\cal P}(f_3,\tau_1;f_4,\tau_2)\cr
&\leq{\cal P}(f_3,\tau_1;-,\tau_2)+{\cal P}(-,\tau_1;f_2,\tau_2)\ .}\eqno(33)
$$
This inequality for the probabilities $\cal P$, involving four generic decay
states $f_i$, $i=1,2,3,4$, of the neutral kaons, is a direct
consequence of the condition (30) and is called a (generalized) Bell's
inequality.

We shall now discuss whether this relation 
is satisfied by ordinary quantum mechanics.
In doing this, one has to compute the probabilities $\cal P$ appearing in (33)
using the expression (13).
Actually, we shall study a simplified version of the inequality, 
obtained from (33)
by setting $f_1=f_4$ and $\tau_1=\tau_2=\tau$. Since the original starting state
$|\psi_A\rangle$ in (8) is antisymmetric, the probability
${\cal P}(f_1,\tau;f_1,\tau)$ identically vanishes in quantum
mechanics, so that (33) reduces to:
$$
\eqalign{
{\cal P}(f_1,\tau;f_2,\tau)
+{\cal P}(f_3,\tau;f_2,\tau)+{\cal P}(f_3,\tau;&f_1,\tau)\cr
&\leq{\cal P}(f_3,\tau;-,\tau)+{\cal P}(-,\tau;f_2,\tau)\ .}\eqno(34)
$$
Furthermore, it is convenient to eliminate the single-kaon probabilities
in the l.h.s. of (34), by using the identity
$$
{\cal P}(f_3,\tau;-,\tau)\equiv 
{\rm Tr}\Big[\Big({\cal O}_{f_3}\otimes{\bf 1}\Big) 
\rho_A(\tau,\tau)\Big]={\cal P}(f_3,\tau;f_1,\tau)+
{\cal P}(f_3,\tau;\tilde f_1,\tau)\ ,\eqno(35)
$$
where ${\bf 1}$ is the $2\times 2$ unit matrix and 
${\cal O}_{f_1}$, ${\cal O}_{\tilde f_1}$ are orthogonal projectors
corresponding to an orthonormal basis in the two-dimensional kaon Hilbert space,
so that ${\bf 1}={\cal O}_{f_1}+{\cal O}_{\tilde f_1}$.
In practice, the use of (35) allows trading the final state $f_1$ for the
different one $\tilde f_1$. In fact,
after these manipulations, the inequality (34) takes the form:
$$
{\cal P}(f_3,\tau;f_2,\tau)-{\cal P}(f_3,\tau;\tilde f_1,\tau)\leq
{\cal P}(\tilde f_1,\tau;f_2,\tau)\ .\eqno(36)
$$
It should be stressed that (36) is a weaker condition than the
original relation (34). Indeed, the relation (35) is compatible with the
expressions (29) for the probabilities $\cal P$ only under the assumption
that the decay of one kaon is stochastically independent from
the decay of the second one.
In other words, in using (35) we assume the absence of variables $\lambda$
that, without violating Bell's locality (30), might nevertheless 
correlate the two decays. This is the price one has to pay for
using an effective two-dimensional description of the $K^0$-$\overline{K^0}$
system. As already observed, if one aims to identify the final states in (36)
with actual asymptotic particles, this is essentially
the only consistent available description. 

The three kaon decay states  
appearing in (36) can be taken to be 
the two-pion or the semileptonic states
discussed before; in particular, for $\tilde f_1=\pi^+\pi^-$,
$f_2=2\pi^0$, $f_3=\pi^-\ell^+\nu$, or $\tilde f_1=2\pi^0$,
$f_2=\pi^+\pi^-$, $f_3=\pi^-\ell^+\nu$, one obtains the two independent
inequalities:
$$
\eqalignno{
&{\cal P}(\pi^-\ell^+\nu,\tau;2\pi^0,\tau)
-{\cal P}(\pi^-\ell^+\nu,\tau;\pi^+\pi^-,\tau)\leq
{\cal P}(\pi^+\pi^-,\tau;2\pi^0,\tau)\ , &(37a)\cr
&{\cal P}(\pi^-\ell^+\nu,\tau;\pi^+\pi^-,\tau)
-{\cal P}(\pi^-\ell^+\nu,\tau;2\pi^0,\tau)\leq
{\cal P}(2\pi^0,\tau;\pi^+\pi^-,\tau)\ ; &(37b)}
$$
since ${\cal P}(f_1,\tau;f_2,\tau)$ is symmetric under the exchange
$f_1\leftrightarrow f_2$, see (13), these can be combined into:
$$
\Big|{\cal P}(\pi^-\ell^+\nu,\tau;2\pi^0,\tau)
-{\cal P}(\pi^-\ell^+\nu,\tau;\pi^+\pi^-,\tau)\Big|\leq
{\cal P}(2\pi^0,\tau;\pi^+\pi^-,\tau)\ .\eqno(38)
$$
A similar relation holds when the semileptonic state $\pi^-\ell^+\nu$
is substituted with $\pi^+\ell^-\bar\nu$.

Explicit expressions for the three probabilities appearing in (38) 
can be easily obtained using the general formula (13) and the normalized
observables ${\cal O}_{+-}$, ${\cal O}_{00}$ and ${\cal O}_{\ell^+}$.
For our purposes, it is sufficient to work with approximate expressions
obtained by keeping only the leading terms in the small 
$CP$ and $CPT$ violating parameters
$\epsilon_L$, $\epsilon_S$, $Y_{+-}$, $Y_{00}$ and 
$\Delta S=\Delta Q$ violating terms $x$ and $z$. 
Up to leading order in all small parameters, one finds:
$$
\eqalignno{
&{\cal P}(\pi^-\ell^+\nu,\tau;2\pi^0,\tau)={1\over4}
e^{-(\gamma_L+\gamma_S)\tau}\Big[1-2\,{\cal R}e(Y_{00})-2\,{\cal R}e(x)\Big]
\ , &(39a)\cr
&{\cal P}(\pi^-\ell^+\nu,\tau;\pi^+\pi^-,\tau)={1\over4}
e^{-(\gamma_L+\gamma_S)\tau}\Big[1-2\,{\cal R}e(Y_{+-})-2\,{\cal R}e(x)\Big]
\ , &(39b)\cr
&{\cal P}(2\pi^0,\tau;\pi^+\pi^-,\tau)={1\over2}
e^{-(\gamma_L+\gamma_S)\tau}\, \big|Y_{+-}-Y_{00}\big|^2\ . &(39c)}
$$
Similar expressions hold for ${\cal P}(\pi^+\ell^-\bar\nu,\tau;2\pi^0,\tau)$
and ${\cal P}(\pi^+\ell^-\bar\nu,\tau;\pi^+\pi^-,\tau)$; they are obtained
from $(39a)$ and $(39b)$ by changing the signs in front of ${\cal R}e(Y_{+-})$ 
and ${\cal R}e(Y_{00})$, and by replacing ${\cal R}e(x)$ with ${\cal R}e(z)$.

Notice that since we are considering probability correlations at equal time,
the time-dependence factorizes. Therefore, inserting (39) in (38), the
time-dependence drops and one finds:
$$
\big|{\cal R}e\big(Y_{+-}-Y_{00}\big)\big|\leq \big|Y_{+-}-Y_{00}\big|^2
\ .\eqno(40)
$$
Recalling the expressions (17), one sees that the difference
$Y_{+-}-Y_{00}$ is expressible solely in terms of the phenomenological
parameter $\varepsilon'$. Then, the assumption of Bell's locality for the system
of correlated neutral kaons coming from the decay of a $\phi$ predicts
that $\varepsilon'$ must satisfy the condition:
$$
\big|{\cal R}e(\varepsilon')\big|\leq 3\, |\varepsilon'|^2\ .\eqno(41)
$$
Therefore, a direct measurement of this parameter in any experimental setup,
not necessarily involving correlated kaons, would automatically be
a check of Bell's locality, within the assumption of stochastic independence
of kaon decays.

The complex parameter $\varepsilon'$ is expected to be very small; 
theoretical estimates based on the Standard Model predict
$|\varepsilon'|\leq 10^{-6}$, while its phase is expected to be
close to $\pi/4$. [13] Therefore, the inequality (41) can be satisfied only
if $\varepsilon'$ is identically zero (such a prediction is made by the
so-called superweak model). The most recent experimental
determinations are not very accurate and the measured value of
${\cal R}e(\varepsilon'/\varepsilon)$, the parameter so far accessible
in actual experiments, does not significantly differ from zero. [15, 16]
However, more refined experimental setups 
are presently under construction, or are just operative, and soon very precise
determination of $\varepsilon'$ will be available.
\footnote{$\!\!\!^\dagger$}{
The fixed target experiments KTEV at Fermilab and NA48 at CERN are already
collecting data; the KLOE apparatus at the DA$\Phi$NE $\phi$-factory in Frascati
is expected to be operative in a year time.}
If these new experiments will confirm the theoretical predictions
for a non-zero $\varepsilon'$, 
than the condition (41) will be violated by several orders of magnitude,
resulting in one of the best tests of Bell's locality.

Before closing, we would like to comment on some of the earlier approaches
to the study of Bell's inequalities in the system of correlated neutral
kaons (see [17-19] and references therein). 
Although the existing literature on the topic is vast, none of these
contributions discuss Bell's locality by taking into account $CP$ and $CPT$
violating effects both in the mass matrix and in the decay amplitudes.
Further, most of the papers deal with a class of Bell's inequalities
which are different from the one considered here; they involve correlated 
decay-probabilities at different times,
not directly connected with asymptotic states, 
{\it i.e.} with experimentally detectable final particles.

The inequalities (36) have also been discussed in Ref.[19], where they have
been derived using different techniques. The aim of that paper is 
to propose a direct experimental test of those inequalities, using a 
$\phi$-factory. Although it is certainly of interest to devise
direct tests of Bell's locality in actual experimental setups, 
we emphasize that our conclusions are independent from
these considerations. Indeed, we were able to reduce the relations (36)
to an equivalent condition on the phenomenological parameter $\varepsilon'$,
directly accessible to the experiment. Any measure of this parameter is
therefore a test on the inequalities (36): there is no need to construct ad hoc
quantum interferometers to check directly these conditions. 
As already observed, such measures
of $\varepsilon'$, besides being crucial for the confirmation
of the Standard Model, will also constitute automatically one of 
the cleanest tests of Bell's locality.

\vskip 2cm

\centerline{\bf ACKNOWLEDGMENTS}
\bigskip

We thank G. Ghirardi, N. Paver and T. Weber for many illuminating discussions.

\vfill\eject

\centerline{\bf REFERENCES}
\bigskip

\item{1.} J. Ellis, J.S. Hagelin, D.V. Nanopoulos and M. Srednicki,
Nucl. Phys. {\bf B241} (1984) 381
\smallskip
\item{2.} J. Ellis, J.L. Lopez, N.E. Mavromatos and D.V. Nanopoulos, 
Phys. Rev. D {\bf 53} (1996) 3846
\smallskip
\item{3.} P. Huet and M.E. Peskin, Nucl. Phys. {\bf B434} (1995) 3
\smallskip
\item{4.} F. Benatti and R. Floreanini, Nucl. Phys. {\bf B488} (1997) 335
\smallskip
\item{5.} F. Benatti and R. Floreanini, Phys. Lett. {\bf B401} (1997) 337
\smallskip
\item{6.} F. Benatti and R. Floreanini, Completely positive dynamics of
correlated neutral kaons, Nucl. Phys. B, to appear
\smallskip
\item{7.} J. Bell, {\it Speakable and Unspeakable in Quantum Mechanics},
(Cambridge University Press, Cambridge, 1987)
\smallskip
\item{8.} J.F. Clauser and M.A. Horne, Phys. Rev. D {\bf 10} (1974) 526
\smallskip
\item{9.} J.F. Clauser and A. Shimony, Rep. Prog. Phys. {\bf 41} (1978) 1881
\smallskip
\item{10.} M. Redhead, {\it Incompleteness, Nonlocality and Realism},
(Clarendon Press, Oxford, 1987)
\smallskip
\item{11.} T.D. Lee and C.S. Wu, Ann. Rev. Nucl. Sci. {\bf 16} (1966) 511
\smallskip
\item{12.} See: C.D. Buchanan, R. Cousins, C. Dib, R.D. Peccei and
J. Quackenbush, Phys. Rev. D {\bf 45} (1992) 4088, and references therein
\smallskip
\item{13.} L. Maiani, in {\it The Second Da$\,\mit\Phi$ne Physics Handbook}, 
L. Maiani, G. Pancheri and N. Paver, eds., (INFN, Frascati, 1995)
\smallskip
\item{14.} N.W. Tanner and R.H. Dalitz, Ann. of Phys. {\bf 171} (1986) 463
\smallskip
\item{15.} NA31-Collaboration, Phys. Lett {\bf B317} (1993) 233
\smallskip
\item{16.} E731-Collaboration, Phys. Rev. D {\bf 55} (1997) 6625
\smallskip
\item{17.} G. Ghirardi, R. Grassi and T. Weber, in the Proceedings
of the {\it Workshop on Physics and Detectors for DA${\mit\Phi}$NE},
G. Pancheri, ed., (INFN-LNF, Frascati, 1991)
\smallskip
\item{18.} P.H. Eberhard, Nucl. Phys. {\bf B398} (1993) 155
\smallskip
\item{19.} A. Di Domenico, Nucl. Phys. {\bf B450} (1995) 293

\bye